\documentclass{PoS}
\usepackage{sidecap}
\usepackage[export]{adjustbox}
\usepackage{capt-of}
\usepackage{caption}

\title{The Present and Future of Searching for Dark Matter with LUX and LZ}

\ShortTitle{Status of LUX and LZ}

\author{\speaker{M. Szydagis}\\
  on behalf of the LUX and LZ collaborations\\
        University at Albany, SUNY\\
        E-mail: \email{mszydagis@albany.edu }}

\abstract{The LUX collaboration new results advance the search for dark matter candidate particles in the 4 $GeV/c^2$ and higher mass range, with a maximal spin-independent 90\% CL limit of $2 \times 10^{-46} cm^2$ at 50 $GeV/c^2$ for its 332 live-day run, following after $6 \times 10^{-46} cm^{2}$ cross-section for 33 $GeV/c^2$ mass from the re-analysis of its initial 95 live-day WIMP search data from December 2015. LUX has performed multiple advanced $in~situ$ neutron and beta/gamma calibrations of light and charge yields down to 1.1 and 0.7 keV, respectively, in nuclear recoil energy and 1.3 and 0.2 keV in units of electron recoil energy, thereby bypassing the past practice of extrapolating yields from $ex~situ$ calibrations or simulation models alone. For this conference proceedings, consequences of the new calibrations for the limit on the interaction cross-sections for low-mass WIMPs will be highlighted. Previous claims of a WIMP signal, from other detectors, are now even more strongly disfavored, assuming isospin invariance and the standard WIMP halo model. Both spin-independent and spin-dependent limits will be discussed, including the recent completion of LUX's 332-live-day blind run. Lastly, we highlight the conceptual design and future plan for its 10-ton-scale, next-generation successor LZ, which plans on achieving $< 3\times10^{-48} cm^2$ sensitivity for a WIMP of mass 40 $GeV/c^2$.}

\FullConference{38th International Conference on High Energy Physics\\
		3-10 August 2016\\
		Chicago, USA}

\begin{document}

\section{Introduction}
Over 80\% of the matter in the universe, or approximately 25\% of its total mass-energy content, continues to elude humanity's scientific community [1]. The two-phase xenon Time Projection Chamber (TPC) has been the world-leading technology over the past few years in the direct search for the WIMP (Weakly Interacting Massive Particle), a leading candidate particle for explaining dark matter [2]. Most recently, this type of device has been exemplified by the LUX (Large Underground Xenon) experiment [3], including its first result from October 2013 [4], and its re-analysis of it from the end of 2015 [5]. LUX was deployed at Sanford Underground Research Facility (SURF), 4850 ft. below the surface of Lead, SD (4300 mwe). LUX was placed in the Raymond Davis Cavern, once home to the Nobel-prize-winning Homestake Experiment which detected solar neutrinos [6]. We will discuss here the details of the internal electric field after the field-generating grid ``conditioning'' campaign, and its 332-live-day WIMP search run thereafter [7], of which 300 live-days were ``salted'' as a form of blinding, with simulated nuclear recoil (NR) events included in the data stream. We will cover the resulting latest LUX sensitivity to the WIMP-nucleon spin-independent (SI) cross-section, the spin-dependent (SD) sensitivities for both neutrons and protons from the first, shorter run [8], and preliminary limits for axions and ALPs (axion-like particles) [9]. Lastly, we will conclude with the future of LUX, and LZ, its multi-ton-scale successor continuing the WIMP quest for a new generation.

In a two-phase Xe TPC, an incoming particle produces scintillation light (S1) in the liquid at 175 nm, which is detected on 10-100 ns timescales by top and bottom arrays of Photomultiplier Tubes (PMTs). The particle will have also ionized atoms. The liberated electrons are drifted upward in an electric field to a gas stage, where they produce their own, O(1 $\mu s$)-wide, scintillation (S2), up to $\sim$300 $\mu s$ after S1. The time in between the S1 and S2 determines the depth of an event in the detector, while the S2 hit pattern in the top PMT array allows reconstruction of the other two dimensions of the event's position. The PMTs convert individual photons into photo-electrons (phe or PE) via the photo-electric effect [10].

Photons detected (phd) is a new unit, created by LUX, that compensates for the $\sim$20\% chance that a single photon produces 2 phe in one of the R8778 Hamamatsu PMTs [11]. LUX additionally uses the digital photon counting method called 'spike counting', at low energy, in order to achieve a better resolution of the S1 signal. This allows for a more precise determination of the integer number of photons initially reaching the PMTs. The internal structure of the LUX detector specifically contains a $\sim$50 x 50 cm dodecagonal cylinder of  PTFE (trade name of Teflon), nearly 100\% reflective at the vacuum ultraviolet (VUV) wavelength of xenon scintillation [12]. The approximate 1:1 ratio of the detector is intended to ensure background radiation does not have a preferred direction for entry into the fiducial volume. LUX contained 370 kg of Xe total, with 250 kg active, here defined as enclosed within the upward drift field. The fiducial mass varied from 118 kg (2013 initial analysis) to 145 kg (2015) to $\sim$100 kg (time- and space-dependent) for the $\sim$1-year data-set. These values depended on the wall background rejection ability. The active volume was instrumented with 122 low-background, VUV-sensitive PMTs, 61 top and bottom. The Xe was pre-purified to remove Kr-85 and other impurities as well as re-circulated during operations, through a commercial hot gas getter, to continuously maintain a high level of purity, for the effective drifting of electrons and transport of photons, and was housed in a low-background titanium cryostat [13]. The energy threshold for NR, that is, the 50\%-efficient point on a sigmoid-like efficiency curve caused by finite energy resolution smearing S1 and S2, was between 3 and 4 keV for the different analyses and runs. This refers to detection of NR prior to the application of electron recoil (ER) background discrimination. In the re-analysis of 2013 data, LUX determined a 0.20\% average ER leakage figure (that is, 99.80\% discrimination) for the WIMP search S1 range of 2 to 50 phd, below the NR log$_{10}$(S2/S1) band per-S1-bin Gaussian centroid. However, this was only a figure of merit since a Profile Likelihood Ratio (PLR) method was used for the limit calculations.

\section{Calibrations}

For both the initial 95 live-day unblinded run and second 332 live-day run, extensive calibrations of LUX were conducted. For NR, this was accomplished with a DD (Deuterium-Deuterium fusion) 2.45 MeV neutron generator source [14]. The neutrons were ``collimated'' by means of an air-filled PVC tube that could be raised into place during calibration, inside of the muon veto water shield. The resulting NR was in the energy range of 0.7-74 keV, making this the most exhaustive calibration of its kind. Through the summing of S1 and S2 per unit energy, excellent agreement was achieved with the Lindhard model, to a degree that for Xe is now comparable to Ge and Si, confirming the hypothesis put forth by Sorensen and Dahl using older data [15]. For ER, the calibration was accomplished with a tritiated methane source (CH$_{3}$T), which could be injected into the Xe, then purified back out with a getter [16]. The resulting ER could be detected in the energy range of 1.3-18.6 keV successfully. This internal calibration avoided reliance on external gamma sources, which find it difficult to penetrate into the fiducial volume of a large-scale TPC LUX-sized [17]. The use of a beta source additionally allowed for a comparison of gamma and beta behavior in Xe at low energies. The primarily Compton background from gamma rays was consistent in S2 and S1 yields with beta data vs. energy [16]. When accounting for the resolution of reconstructed energies combining S1 and S2, the beta spectrum could be reconstructed, and a fall-off from threshold at low energy reproduced in simulation [18]. Fig. 1 shows the source delivery systems for both calibrations.

\begin{SCfigure}
  \includegraphics[width=8cm]{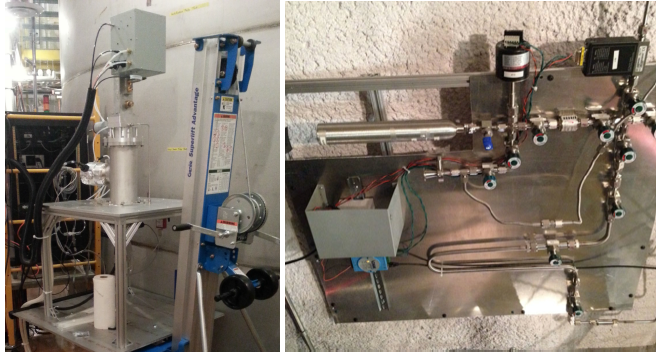}
 \caption{\small{Left: The DD neutron ``gun'' [38] behind poly shielding. It resided immediately next to the water tank, and emitted n's into 4$\pi$, with neutrons directed using a pipe inside to maintain beam purity, in terms of original energy. Right: Delivery system into the Xe stream for the methane, where one ordinary hydrogen atom was replaced with tritium. Info on data taken using these sources is within [14,16].}}
  \label{fig1}
\end{SCfigure}

\section{Understanding the New Data}

The drift electric field was 180 V/cm and extraction field was 6 kV/cm during the first science run. Both fields were successfully raised for the subsequent run, after a period of ``conditioning.'' This was motivated by expected higher ER discrimination at higher electric fields [19], though S1 photon detection efficiency turned out to be more important [20]. It was also motivated by a desire to lower the S2 threshold, with an improved extraction of electrons into the gas amplification region, which increased from 48.9\% to over 70\%. However, the drift field became distorted, both in magnitude, between 50 and 600 V/cm, and in direction, with radial components distorting the position reconstruction and thus the determination of the fiducial volume boundary. The fields continued to change over the course of the run. These issues were addressed by splitting up the data from the 332 live-day run into 16 bins, in time, based on the dates of events, and in space, using the drift time as the standard indicator of event depth. There were 4 slices in time of occurrence and 4 in space (in depth, known as ``Z-slices''), to make up these 16 independent detectors essentially, each with its own electric field and own S1 and S2 efficiencies. The number of (equally-sized) spatial slices were chosen so that the ER and NR band Gaussian means as a function of S1 would not vary significantly within their uncertainties inside of them; the time slices, which were unequal, and spanned 2014 to 2016, were chosen based on when the slow drift over time of the reconstructed radius to the wall exceeded the error on it, due to the time-varying field. The fiducial mass for each time/space bin varied between 98-105 kg, chosen to minimize backgrounds from the wall.

The NEST (Noble Element Simulation Technique) ever-updating, semi-empirical umbrella of light and charge yield models [21] was utilized to construct a series of Monte Carlo simulations, for NR and for ER, for each of the 16 segments. The formulae and results presented in [22] and [23] respectively for ER and NR were updated with slightly modified, newer values, based on the tritium and DD calibrations of the first LUX WIMP search, providing the energy dependence of the ionization and light yields. The field dependencies of these yields were provided by studying PandaX small-scale chamber calibration data [24]. The same LUX calibrations were continued and repeated for the second science run, but were not treated as input into simulation. The simulation assumed yields from the existing NEST version were correct, adjusting S1 and S2 efficiencies to compensate, and allowing the mean field to float as a free parameter, comparing it to independent measurements of field. The result was that the best-fit electric fields from NEST agreed within uncertainty, statistical as well as the systematic from radial variation in field, with the COMSOL modeling of the time- and space-varying field. This established the reliability of simulation in terms of its service as a basis for PLR signal and background models (Fig. 2).

\begin{SCfigure}
  \includegraphics[width=8cm]{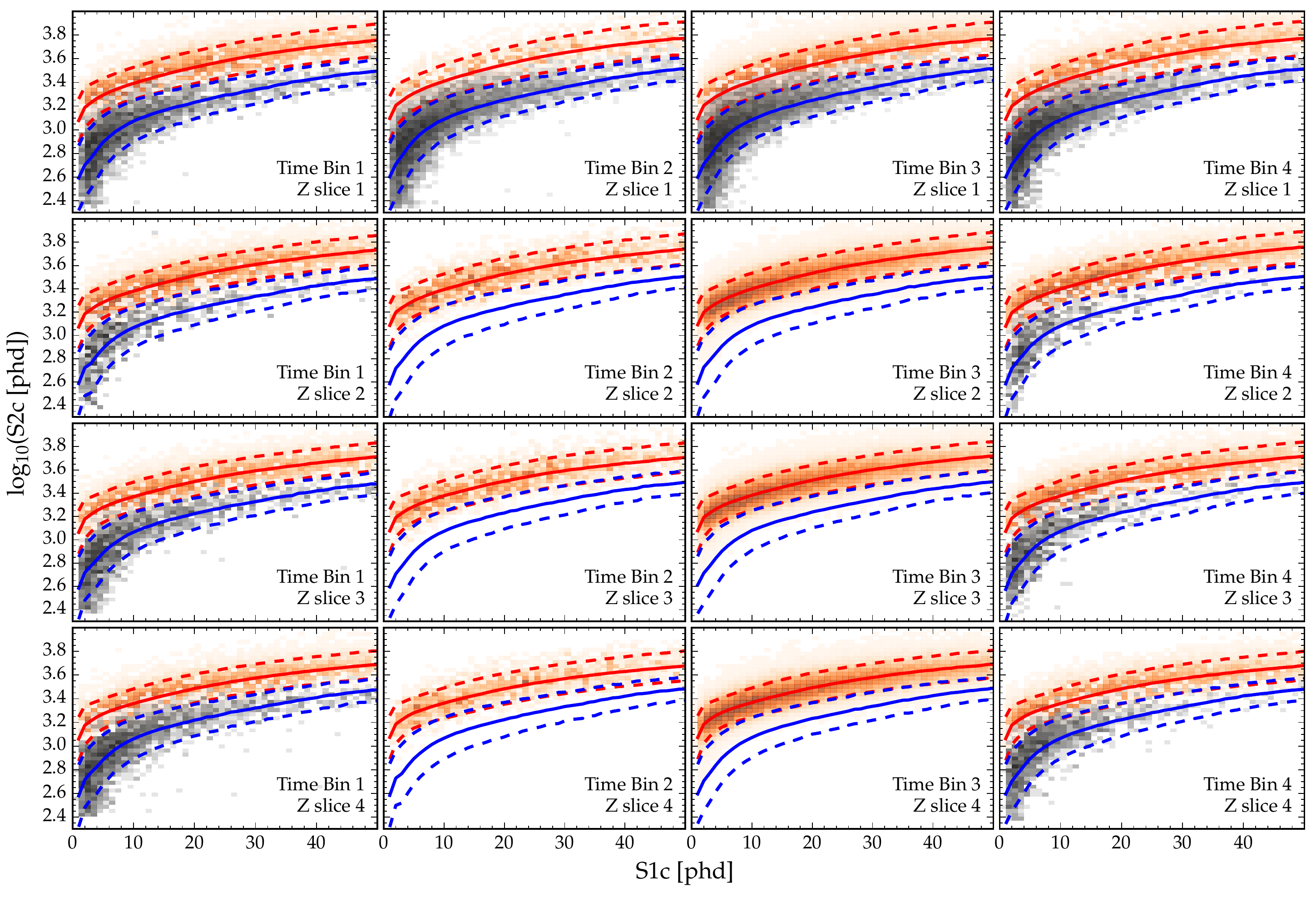}
  \caption{\small{The time bins and Z-slices of the LUX detector for its second, longer run, when the electric drift field became distorted. The gray densities represent the CH$_{3}$T calibrations (ER) and the orange densities are the DD calibrations (NR). The solid lines are the NEST model band means, tuned, while the dashes are 10-90 percentiles. Not every bin or slice has both calibration types with good statistics, but the good agreement between NEST and data allowed interpolation. Top row represents top of the detector. Left to right is 9/2014 through 5/2016.}}
  \label{fig2}
\end{SCfigure}

Prior to un-salting and application of a 2-sided, un-binned PLR, the
background model was verified through comparison with data. The upper
half of the ER band was used a side band, the lower half having been
contaminated on purpose with fake WIMP signals. Comparison of the
background data was done in the five-dimensional space of the PLR:
event radius, angle, drift time, S1, and log(S2), the first three
defining position and latter two energy. Figure 3 shows simulation and
data for each of the five. The KS test p-values in each dimension far
exceeded 10\% typically, equal to or greater than the previous run's
well-fitting background model [25]. The number of events observed
below the NR band Gaussian average was compared to the expected
numbers during un-salting, broken down by category in Table 1.


\begin{figure}[htb]
\begin{minipage}[b]{0.5\linewidth}
\centering
\includegraphics[width=7cm]{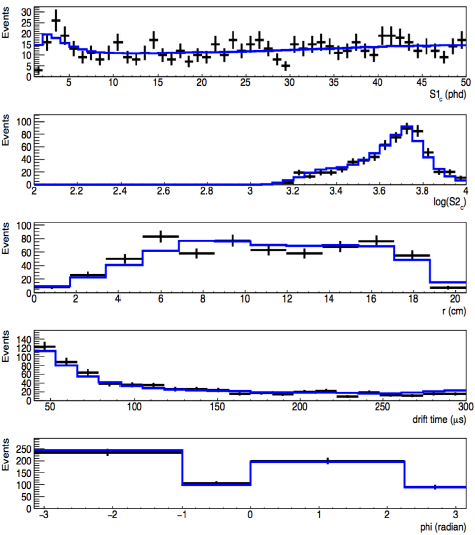}
\captionsetup{}
\caption{\footnotesize{Comparison of data (black) to sim (blue) for the ER background in the upper half of the ER band, across the five dimensions of the PLR, as part of the pre-un-salting verifications.}}
\end{minipage}%
\hspace{0.2cm}%
\begin{minipage}[b]{0.5\linewidth}
\centering
\small
\begin{tabular}{|c|c|}
\hline
Background Source             & \# Expected     \\ 
\hline
External gamma rays (all)     & 1.51 $\pm$ 0.19   \\
Internal betas                & 1.20 $\pm$ 0.06    \\
Radon plate out (wall BG)     & 8.7 $\pm$ 3.5     \\
Accidental S1-S2 coincidences & 0.34 $\pm$ 0.10    \\
Solar $^{8}$B neutrinos (CNNS) & 0.15 $\pm$ 0.02   \\
Neutrons from all sources     & $\sim$O(0.1) \\ 
\hline
\end{tabular}
\captionsetup{width=7cm}
\captionof{table}{\footnotesize{The expected number of sub-NR-mean events, for all backgrounds, broken down into broad categories, used only as a figure of merit. Gammas/betas are in the bulk volume, with leakage at all energies. Rn is low-energy but confined to the edge of the volume; the PLR includes position information, so these events have low signal likelihood despite there being many of them. The last two are in the bulk volume, low-energy, and within the NR band itself, but each <1 event in expectation.} }
\label{table1}
\end{minipage}
\end{figure}

\section{Dark Matter Search Results}

Traditional blinding masks the signal region, the NR band, completely,
and more, the lower half of the ER band as well, in order to account
for statistical fluctuations upward. The challenge often seen in the
direct detection community is the side effect of blindness to rare
backgrounds and pathologies [26]. One may not need to go to such great
lengths to mitigate the potential bias. Instead of traditional
blinding, LUX employs a technique where fake signal events, called
``salt'' are injected into the data stream. Salt uses simulation only to
ensure a proper S2/S1 ratio, drawing from several differential
exponentials and a uniform-in-energy distribution as well, not
representing any particular WIMP mass. Real data from calibration
(tritium, which was higher-statistics than DD, which needed to be kept
for the critical NR calibration) are used for the pulse selection and
pairing. Salt mitigates bias while allowing for scrutiny of individual
events and has been already used to great effect in neutrino
experiments and searches for free particles with fractional charge
[27]. The salting was performed successfully, with the fake events
being all removed at the end, but prior to the limit calculation
events outside of the ER band were re-scrutinized. Two populations of
rare, pathological events were identified, but only after the ``desalination'' process, contributing 3 events below the NR band
mean. In a likelihood analysis, events such as these which match no
background distribution pose more of a worry than they would in a cut
and count analysis where they may be absorbed as a statistical
fluctuation. It was discovered that S1 quality cuts had simply been
lacking, since single electron backgrounds had made the focus be the
S2 quality cuts. Post unblinding cuts were thus created, targeting gas
S1 events and Cerenkov-like events (light mostly in 1 PMT), both
clearly not WIMP-like. These cuts were soft, independent of energy and
high in signal acceptance (NR), and defined using only the large
calibration data sets, not tuned specifically to the 3 events in WIMP
search.

	The p-value for the background-only model was 40\%, after the
        removal of the salt events, plus the cuts removing the small
        subset of events strongly discrepant with the vast majority of
        calibration events based on different S1 aspects. The
        resulting WIMP-nucleon SI exclusion (90\% C.L.) versus mass
        can be seen in Fig. 4, with the best, lowest exclusion at 50
        GeV, of $2.2 \times 10^{-46} cm^2$ or 0.22 zeptobarns (zb) in
        cross-section. This is within 1 order of magnitude of the
        XENON1T projection [28], and within 2 of LZ [29]. Fig. 4 also
        includes the LUX re-analysis (2015) of the 2013 data (labeled
        LUX 2015). The new exclusion curve is comparable to the LUX
        2015 re-analysis of 3 months' worth of data at low mass,
        despite nearly 3.5x more live-time, due to an upward
        statistical fluctuation in the number of background events at
        the lowest energies (still well within 1-2 sigma). This
        re-analysis however, was over 2-3 orders of magnitude better
        at low WIMP masses compared to the initial LUX 2014 curve,
        which itself was already in conflict with the results from
        DAMA [30] and CoGeNT [31]. That improvement was driven by the
        new DD calibrations, which demonstrated NR light and charge
        yields below 3 keV (down to sub-keV), where a conservative
        cut-off had been set, assuming no yield below it. At high mass
        the limit is 4x better, slightly better than one would assume
        from increase in exposure, due to a favorable, downward
        fluctuation, at the highest energies of the WIMP search data
        (1-sigma below expectation). The 332 live days of data best
        the 95 live-day result minimum cross-section in the limit of
        0.60 zb at 33 GeV (Fig 4., which also contains projections for
        1,000 live-days of LZ running, default planned, starting in
        2020 [29]).

	The SD results appear in [8], which used only the initial 95 live-day data. These results will be updated in the near future to include the data from the 332 live-day exposure. Xe remains the best element to use for SD-neutron coupling. Preliminary results from LUX for both solar and galactic axions/ALPs in terms of coupling to electrons can be found in [9], using only the first 95 live days, with exclusion slightly better than that from XENON100, a similar detector [32].

\begin{SCfigure}
\centering
  \includegraphics[width=8cm]{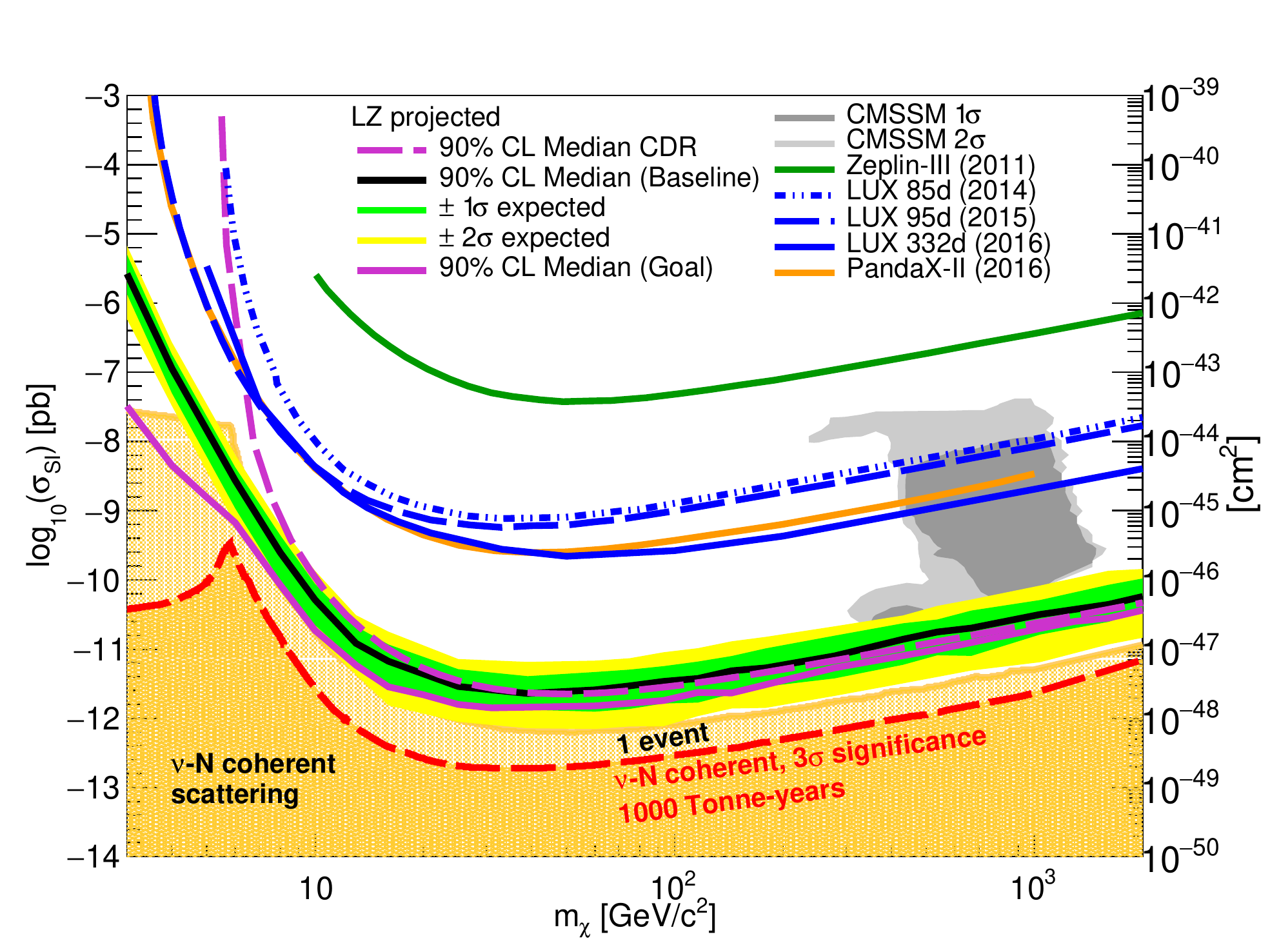}
  \caption{\footnotesize{All 3 LUX analyses, 2014, 2015, and 2016, with 2 different data sets (2014 and 2015, vs. 2016, the current result). ZEPLIN-III final from 2011 is included [33], as well as example SUSY models [34]. Different shades of orange at bottom indicate different degrees of interference from the neutrino background from the sun and other sources. The main goal of LZ is $< 3\times 10^{-48} cm^2$ at 40 GeV. Between CDR (Conceptual Design Report, pink dashed) and new baseline for TDR (Technical Design Report) the 3 keV cut was dropped (black); best case 'goal' added (pink solid). The baseline involves the conservative meeting of the minimum requirements [29].}}
  \label{fig4}
\end{SCfigure}

\section{LZ and Conclusion}

The LUX-ZEPLIN (LZ) collaboration is a merger of the two former Xe TPC competitors, a larger as well as improved version compared to either LUX or ZEPLIN chosen by the DOE G2 down-select. LZ has passed its CD-1 (Critical Decision) and CD-2 reviews successfully and is only awaiting final CD-3 approval in 2017. Construction on certain detector elements has been underway since 2015, and all elements will commence construction in early 2017. LZ will use existing SURF infrastructure from the decommissioned LUX, especially the water tank. The total mass of xenon will be 10 tonnes , with 7 active and $\sim$5.6 in the fiducial volume. LZ will possess a unique triple veto system, composed of a PMT-instrumented (S1 only) Xe skin primarily serving as active gamma shielding, a Gd-loaded liquid scintillator for neutron tagging, and a Cerenkov muon veto for cosmic rays. Inclusion of these vetos increases the fiducial mass estimate from an original $\sim$3 tonnes: Xe used previously only for self-shielding of backgrounds is now ``recovered.'' These vetos also increase confidence for discovery. The conservative baseline requirements of LZ include a 6 keVnr threshold, with at least 99.5\% discrimination (LUX and ZEPLIN have both already demonstrated better for each). Many Higgs-mediated models will be probed by LZ, including a favored 1 TeV Higgsino [35].

In summary, world-leading results, from LUX's 332 and 95 live-day searches for dark matter have cut significantly deeper into previously un-probed parameter space. The latest, final run has an exposure of 33,500 kg-days, the most of any 2-phase Xe TPC to date. PandaX is close, with a much larger mass but less live-time [36], so it will likely exceed this exposure soon, followed by the even larger XENON1T [28]. More publications will be forthcoming soon, including the combination of the results from the two runs into one, combined LUX exclusion plot [7]. Again at SURF, LZ will deploy a multi-ton-scale detector as one of only 3 (2 for WIMPs, 1 for axions) down-selected G2 experiments, and come within 1 order of magnitude of the coherent scattering neutrino floor at high mass, ``clipping'' the neutrino ``shoulder'' at low [37]. For a 40 GeV WIMP just outside the constraints of the current limits, it is anticipated to see 3$\sigma$ evidence. LZ should have the best detector with hope for a WIMP discovery soon.

\section{Acknowledgments}

This work was partially supported by the U.S. Department of Energy (DOE) under award numbers DE-SC0012704, DE-AC02-05CH11231, DE-SC0012161, DE-SC0014223, DE-FG02-13ER42020, DE-FG02-91ER40674, DE-NA0000979, DE-SC0011702, DESC0006572, DESC0012034, DE-SC0006605, DE-FG02-10ER46709, DE-AC05-06OR23100, DE-AC52-07NA27344, DE-FG01-91ER40618, DEFG02-08ER41549, DE-FG02-11ER41738, DE-FG02-91ER40688, DE-FG02-95ER40917, DESC0010010, and DE-SC0015535; by the U.S. National Science Foundation (NSF) under award numbers NSF PHY-110447, NSF PHY-1506068, NSF PHY-1312561, NSF PHY-1406943, NSF PHY-0750671, NSF PHY-0801536, NSF PHY-1003660, NSF PHY-1004661, NSF PHY-1102470, NSF PHY-1347449, NSF PHY-1505868, and NSF PHY-1636738; the Research Corporation grant RA0350; the South Dakota Governor's Research Center for the Center for Ultra-low Background Experiments in the Dakotas (CUBED); the South Dakota School of Mines and Technology (SDSMT); and from the University of Wisconsin for grant UW PRJ82AJ. LIP-Coimbra acknowledges funding from Funda\c{c}$\tilde{a}$o para
a Ci$\hat{e}$ncia e a Tecnologia (FCT) through the project grant $PTDC/FIS-NUC/1525/2014$. Imperial College and Brown University thank the UK Royal Society for travel funds under the International Exchange Scheme (IE120804). The UK groups acknowledge institutional support from Imperial College London, University College London, and Edinburgh University, and from the UK Science \& Technology Facilities Council for Ph.D. studentships under award numbers ST/K006428/1, ST/M003655/1, ST/M003981/1, ST/M003744/1, ST/M003639/1, ST/M003604/1, ST/M003469/1, $ST/K502042/1 (AB)$, $ST/K502406/1 (SS)$, and $ST/M503538/1 (KY$). The University of Edinburgh is a charitable body, registered in Scotland, with registration number SC005336.

This research was conducted using computational resources and services at the Center for Computation and Visualization, Brown University, and also the Yale Science Research Software Core. The $^{83}$Rb used in this research to produce $^{83m}$Kr was supplied by the United States Department of Energy Office of Science by the Isotope Program in the Office of Nuclear Physics. We gratefully acknowledge the many types of logistical and technical support and the access to laboratory infrastructure provided to us by the Sanford Underground Research Facility (SURF) and its personnel at Lead, South Dakota. SURF was developed by the South Dakota Science and Technology Authority (SDSTA), with an important philanthropic donation from T. Denny Sanford as well as support from the State of South Dakota, and is operated by Lawrence Berkeley National Laboratory for the Department of Energy, Office of High Energy Physics.

\section{References}

[1] G.R. Blumenthal, S. Faber, J.R. Primack, and M.J. Rees, ``Formation of Galaxies and Large-scale Structure with Cold Dark Matter,'' Nature 311, 517 (1984); J.L. Feng, ``Dark Matter Candidates from Particle Physics and Methods of Detection,'' Ann. Rev. Astron. Astroph. 48, 495 (2010). arXiv:1003.0904

[2] J.D. Lewin and P.F. Smith, `` Review of Mathematics, Numerical Factors, and Corrections for Dark Matter Experiments Based on Elastic Nuclear Recoil,'' Astropart. Phys. 6, 87-112 (1996).8

[3] D.S. Akerib et al. (LUX Collaboration), `` The Large Underground Xenon (LUX) Experiment,'' Nucl. Inst. Meth. A704, 111 - 126 (2013). arXiv:1211.3788

[4] D.S. Akerib et al., `` First Results from the LUX Dark Matter Experiment at the Sanford Underground Research Facility,'' Phys. Rev. Lett. 112, 091303 (2014). arXiv:1310.8214

[5] D.S. Akerib et al., ``Improved Limits on Scattering of Weakly Interacting Massive Particles from Re-analysis of 2013 LUX Data,'' Phys. Rev. Lett 116, 161301 (2016). arXiv:1512.03506

[6] K.T. Lesko et al., `` Deep Underground Science and Engineering Laboratory - Preliminary Design Report'', arXiv:1108.0959 [hep-ex]; K.T. Lesko, ``The Sanford Underground Research Facility at Homestake,'' Eur. Phys. J. Plus 127, 107 (2012); D.-M. Mei, C. Zhang, K. Thomas, and F.E. Gray, ``Early Results on Radioactive Background Characterization for Sanford Laboratory and DUSEL Experiments,'' Astropart. Phys. 34, 33-39 (2010); F.E. Gray, C. Ruybal, J. Totushek, D.-M. Mei, K. Thomas, and C. Zhang, ``Cosmic Ray Muon Flux at the Sanford Underground Laboratory (SUL) at Homestake,'' Nucl. Inst. Meth. A638, 63 (2011). arXiv:1007.1921

[7] D.S. Akerib et al., ``Results from a Search for Dark Matter in the Complete LUX Exposure,'' Submitted to Phys. Rev. Lett. (2016). arXiv:1608.07648

[8] D.S. Akerib et al., ``Results on Spin-Dependent Scattering of Weakly Interacting Massive Particles (WIMPs) on Nucleons in the Run 3 Data of the LUX Experiment,'' Phys. Rev. Lett. 116, 161302 (2016). arXiv:1602.03489

[9] M.F. Marzioni et al. (LUX Collaboration), ``Axion and Axion-Like Particle Searches in LUX,'' IDM conference presentation (July 2016).
 
https://idm2016.shef.ac.uk/indico/event/0/session/6/contribution/95

[10] B.A. Dolgoshein, V.N. Lebedenko, and B.U. Rodionov, ``New Method of Registration of Ionizing- particle Tracks in Condensed Matter,'' JETP Lett. 11, 513 (1970); D. Nygren, ``Optimal Detectors for WIMP and $0\nu\beta\beta$ Searches: Identical High-Pressure Xenon Gas TPCs?'' Nucl. Inst. Meth. A581, 632-642 (2007).

[11] C.H. Faham, V.M. Gehman, A. Currie, A. Dobi, P. Sorensen, and R.J. Gaitskell, ``Measurements of Wavelength-dependent Double Photoelectron Emission from Single Photons in VUV-sensitive Photomultiplier Tubes,'' JINST 10, P09010 (2015). arXiv:1506.08748

[12] C. Silva, J. Pinto da Cunha, A. Pereira, M.I. Lopes, V. Chepel, V. Solovov, and F. Neves, ``A Model of the Reflection Distribution in the Vacuum Ultraviolet Region,'' Nucl. Inst. Meth. A619, 59-62 (2010). arXiv:0910.1058; C. Silva, J. Pinto da Cunha, A. Pereira, V. Chepel, M.I. Lopes, and V. Solovov, ``Reflectance of Polytetrafluoroethylene (PTFE) for Xenon Scintillation Light,'' arXiv:0910.1056 

[13] D.S. Akerib et al., ``Radio-assay of Titanium Samples for the LUX Experiment,'' 

arXiv:1112.1376 

[14] D.S. Akerib et al., `` Low-energy (0.7-7.4 keV) Nuclear Recoil Calibration of the LUX Dark Matter Experiment Using D-D Neutron Scattering Kinematics,'' Submitted to Phys. Rev. C (2016). arXiv:1608.05381

[15] P. Sorensen and C.E. Dahl, `` Nuclear recoil energy scale in liquid xenon with application to the direct detection of dark matter,'' Phys. Rev. D 83, 063501 (2011). arXiv:1101.6080

[16] D.S. Akerib et al., ``Tritium Calibration of the LUX Dark Matter Experiment,'' Phys. Rev. D 93, 072009 (2016). arXiv:1512.03133

[17] E. Aprile et al. (XENON100 Collaboration), ``Dark Matter Results from 225 Live Days of XENON100 Data,'' Phys. Rev. Lett. 109, 181301 (2012). arXiv:1207.5988

[18] D.S. Akerib et al., ``LUXSim: A Component-Centric Approach to Low-Background Simulations,'' Nucl. Inst. Meth. A675, 63 (2012). arXiv:1111.2074

[19] V.N. Lebedenko et al. (ZEPLIN-III Collaboration), ``Results from the First Science Run of the ZEPLIN-III Dark Matter Search Experiment,'' Phys. Rev. D 80, 052010 (2009). arXiv:0812.1150 

[20] C.E. Dahl, ``The Physics of Background Discrimination in Liquid Xenon, and First Results from XENON10 in the Hunt for WIMP Dark Matter,'' Ph.D. thesis, Princeton University, January 2009.

[21] M. Szydagis, N. Barry, K. Kazkaz, J. Mock, D. Stolp, M. Sweany, M. Tripathi, S. Uvarov, N. Walsh, and M. Woods, ``NEST: A Comprehensive Model for Scintillation Yield in Liquid Xenon,'' JINST 6, P10002 (2011). arXiv:1106.1613

[22] M. Szydagis, A. Fyhrie, D. Thorngren, and M. Tripathi, ``Enhancement of NEST Capabilities for Simulating Low-Energy Recoils in Liquid Xenon,'' JINST 8, C10003 (2013). arXiv:1307.6601 

[23] B. Lenardo, K. Kazkaz, A. Manalaysay, J. Mock, M. Szydagis, and M. Tripathi, ``A Global Analysis of Light and Charge Yields in Liquid Xenon,'' IEEE Trans. Nucl. Sci. 62, 3387 (2015). arXiv:1412.4417

[24] Q. Lin, J. Fei, F. Gao, J. Hu, Y. Wei, X. Xiao, H. Wang, and K. Ni, ``Scintillation and Ionization Responses of Liquid Xenon to Low Energy Electronic and Nuclear Recoils at Drift Fields from 236 V/cm to 3.93 kV/cm,'' Phys. Rev. D 92, 032005 (2015). arXiv:1505.00517

[25] D.S. Akerib et al., ``Radiogenic and Muon-Induced Backgrounds in the LUX Dark Matter Detector'', Astropart. Phys. 62, 33 - 46 (2015). arXiv:1403.1299

[26] E. Aprile et al. (XENON100 Collaboration), `` Analysis of the XENON100 Dark Matter Search Data,'' Astropart. Phys. 54, 11-24 (2014). arXiv:1207.3458

[27] A. Giachero et al. (CUORE Collaboration), `` The CUORE and CUORE-0 Experiments at Gran Sasso,'' Proceedings of a talk given at the ICNFP 2014 conference, submitted at EPJ Web of Conferences (2014). arXiv:1410.7481

[28] E. Aprile et al. (XENON1T Collaboration), ``Physics Reach of the XENON1T Dark Matter Experiment,'' JCAP 04, 027 (2016). arXiv:1512.07501

[29] D.S. Akerib et al. (LZ Collaboration), ``LUX-ZEPLIN (LZ) Conceptual Design Report,'' U.S. Department of Energy (DoE) technical paper, 2015. arXiv:1509.02910 (arXiv only)

[30] R. Bernabei et al. (DAMA Collaboration), ``First Results from DAMA/LIBRA and the Combined Results with DAMA/NaI,'' Eur. Phys. J. C56, 333 (2008). arXiv:0804.2741 

[31] C.E. Aalseth et al. (CoGeNT Collaboration), ``CoGeNT: A Search for Low-Mass Dark Matter using p-type Point Contact Germanium Detectors,'' Phys. Rev. D88 No. 1, 012002 (2013). arXiv:1208.5737 

[32] E. Aprile et al. (XENON100 Collaboration), ``First Axion Results from the XENON100 Experiment,'' Phys. Rev. D 90, 062009 (2014). arXiv:1404.1455

[33] D.Yu. Akimov et al. (ZEPLIN-III Collaboration), ``WIMP-Nucleon Cross-Section Results from the Second Science Run of ZEPLIN-III,'' Phys. Lett. B 709, 14-20 (2012). arXiv:1110.4769

[34] E.A. Bagnaschi et al., ``Supersymmetric Dark Matter after LHC Run 1,'' Eur. Phys. J. C75, 500 (2015). arXiv:1508.01173

[35] L. Roszkowski, E.M. Sessolo, and A.J. Williams, ``Prospects for Dark Matter Searches in the pMSSM,'' JHEP 1502, 014 (2015). arXiv:1411.5214

[36] A. Tan et al. (PandaX-II Collaboration), ``Dark Matter Results from First 98.7-day Data of PandaX-II Experiment,'' Phys. Rev. Lett. 117, 121303 (2016). arXiv:1607.07400

[37] J. Billard, E. Figueroa-Feliciano, and L. Strigari, ``Implication of Neutrino Backgrounds on the Reach of Next Generation Dark Matter Direct Detection Experiments,'' Phys. Rev. D 89, 023524 (2014). arXiv:1307.5458

[38] J. Verbus et al., `` Proposed Low-energy Absolute Calibration of Nuclear Recoils in a Dual-Phase Noble Element TPC Using D-D Neutron Scattering Kinematics.'' arXiv:1608.05309



\end{document}